\documentclass[3p,times]{elsarticle}

\usepackage{ecrc}

\pdfmapfile{+txfonts.map}

\usepackage{epstopdf}

\volume{00}

\firstpage{1}

\journalname{Nuclear Physics A}

\runauth{Joseph I. Kapusta}

\jid{nupha}

\jnltitlelogo{Nuclear Physics A}

\usepackage{graphicx}
\usepackage{amsmath,amssymb}

\newcommand{\be}{\begin{equation}}
\newcommand{\ee}{\end{equation}}
\newcommand{\ba}{\begin{eqnarray}}
\newcommand{\ea}{\end{eqnarray}}
\newcommand{\bd}{\begin{displaymath}}
\newcommand{\ed}{\end{displaymath}}

\begin{document}

\begin{frontmatter}



\title{High Baryon Densities Achieveable at RHIC and LHC}

\author{Joseph I. Kapusta and Ming Li}
\address{School of Physics and Astronomy, University of Minnesota,
Minneapolis MN 55455, USA}




\begin{abstract}
In high energy collisions nuclei are practically transparent to each other but produce very hot, nearly baryon-free, matter in the central rapidity region. Where do the baryons go? We calculate the energy loss of the nuclei using the color glass condensate model. Using a space-time picture of the collision we calculate the baryon and energy densities of the receding baryonic fireballs. For central collisions of large nuclei at RHIC and LHC we find baryon densities more than ten times that of atomic nuclei over a large volume which appear at high rapidities. These results can and are being used as initial conditions for subsequent hydrodynamic evolution and could test the equation of state of matter at very high baryon densities.
\end{abstract}


\end{frontmatter}



\section{Introduction}

For thirty-five years the high energy heavy ion community has been focussed on the central rapidity region following the seminal work of Bjorken \cite{BJ1983}.  The reasons are (i) the energy density is expected to be higher there, (ii) the matter is nearly baryon-free making it relevant to the type of matter that existed in the early universe and more amenable to comparisons with lattice gauge calculations, and (iii) detectors in a collider can more readily measure particle production and correlations within one unit of rapidity around the center-of-momentum.  The preceding work of Anishetty, Koehler and McLerran \cite{AKM1980}, which found that nuclei were compressed by a factor of 3.5 and excited to an energy density of about 2 GeV/fm$^3$ when they collide at extreme relativistic energies, was pursued only sporadically.  Here we report on our recent work on the baryonic fireballs which emerge beyond one unit of rapidity \cite{LKRapid}.  Our calculations are not particularly relevant to the lower energy beam scans at RHIC.

\section{Nuclear Rapidity Loss and Internal Excitation}

In the color glass condensate picture the color charges in the nuclei are essentially frozen during the very short time that it takes for the nuclei to pass through each other \cite{McLerranVenu}.  These charges are the source for the initial fields at $\tau=0$ which provide the initial conditions for solving the classical Yang-Mills equations.  The early time development of the so-called glasma is represented by the energy-momentum tensor of the form \cite{CFKL2015,LK2016}
\begin{equation}\label{em_tensor}
T^{\mu\nu}_{\rm glasma}=
\begin{pmatrix}
{\cal A}+{\cal B}\cosh{2\eta} & 0 & 0 & {\cal B}\sinh{2\eta} \\
0 & {\cal A} & 0 & 0 \\
0 & 0 & {\cal A} & 0 \\
{\cal B}\sinh{2\eta} & 0 & 0 & -{\cal A}+{\cal B}\cosh{2\eta} \\
\end{pmatrix} \, .
\end{equation}
The $\mathcal{A}$ and $\mathcal{B}$ are known analytical \cite{LK2016} functions of proper time $\tau$, while the dependence on space-time rapidity $\eta$ follows from the fact that $T^{\mu\nu}_{\rm glasma}$ is a second-rank tensor in a boost-invariant setting.  The longitudinal position $z_{\rm P}$ at each ${\bf r}_\perp$ is a function of time.  The $z_{\rm P}(t)$ is related to the time $t$ via the velocity $v_{\rm P} = dz_{\rm P}/dt =\tanh{y_{\rm P}}$, where $y_{\rm P}$ is the momentum-space rapidity of the part of the nucleus located at ${\bf r}_\perp$. The Lorentz invariant effective mass per unit area ${\cal M}_{\rm P}$ is defined via the relations ${\cal E}_{\rm P} = {\cal M}_{\rm P} \cosh{y_{\rm P}}$ and ${\cal P}_{\rm P} = {\cal M}_{\rm P} \sinh{y_{\rm P}}$.  The pair of differential equations 
\begin{equation}\label{eom}
\begin{split}
&d{\cal E}_{\rm P}(t,z_{\rm P}) = -T^{00}_{\rm glasma}(t,z_{\rm P})dz_{\rm P} + T^{03}_{\rm glasma}(t,z_{\rm P})dt \\
&d{\cal P}_{\rm P}(t,z_{\rm P}) = -T^{30}_{\rm glasma}(t,z_{\rm P})dz_{\rm P} + T^{33}_{\rm glasma}(t,z_{\rm P})dt \,. \\
\end{split}
\end{equation}
describe both the loss of kinetic energy of the projectile nucleus and the internal excitation energy imparted to it during the collision.  Thus ${\cal M}_{\rm P}$ is not constant but increases with time, unlike the case of the string model \cite{MK2002}.  This is a very important difference.  Nuclear matter cannot be compressed without providing the necessary energy.

Several scales enter the problem.  Perhaps the most interesting one is a transverse UV regulator $Q$, which is the scale separating the soft physics represented by the classical gluon fields and the hard physics represented by jets and mini-jets.  When the contribution to the energy-momentum tensor from jets and mini-jets is included, which we are not doing here, the sum total should be relatively insensitive to the prescise choice of $Q$.  In Fig. 1 we show the rapidity of the central core of a Au projectile as a function of proper time.  It loses nearly 3 units of rapidity in the first 0.1 fm/c with little sensitivity to $Q$.  In Fig. 2 we show the excitation energy per baryon of the same central core.  Most of the excitation energy is gained between 0.05 and 0.5 fm/c with a relatively mild sensitivity to $Q$. 
\begin{figure}[h]
\begin{minipage}{18pc}
\includegraphics[width=18pc]{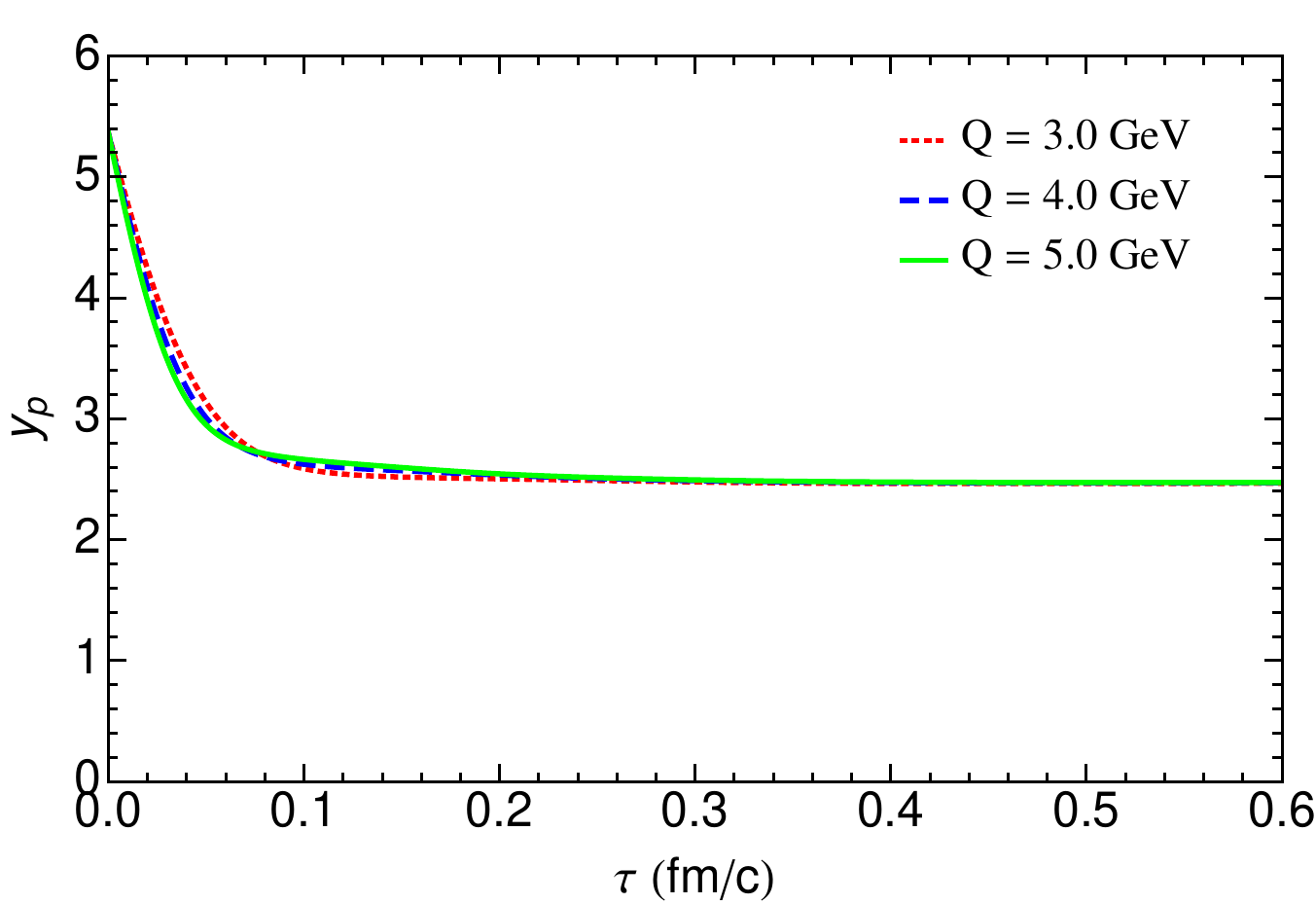}
\caption{\label{F1} Rapidity of the central core of a Au projectile nucleus in the center-of-momentum frame for $\sqrt{s_{NN}} = 200$ GeV as a function of proper time.  The result is insensitive to the choice of $Q$.}
\end{minipage}\hspace{2pc}%
\begin{minipage}{18pc}
\includegraphics[width=18pc]{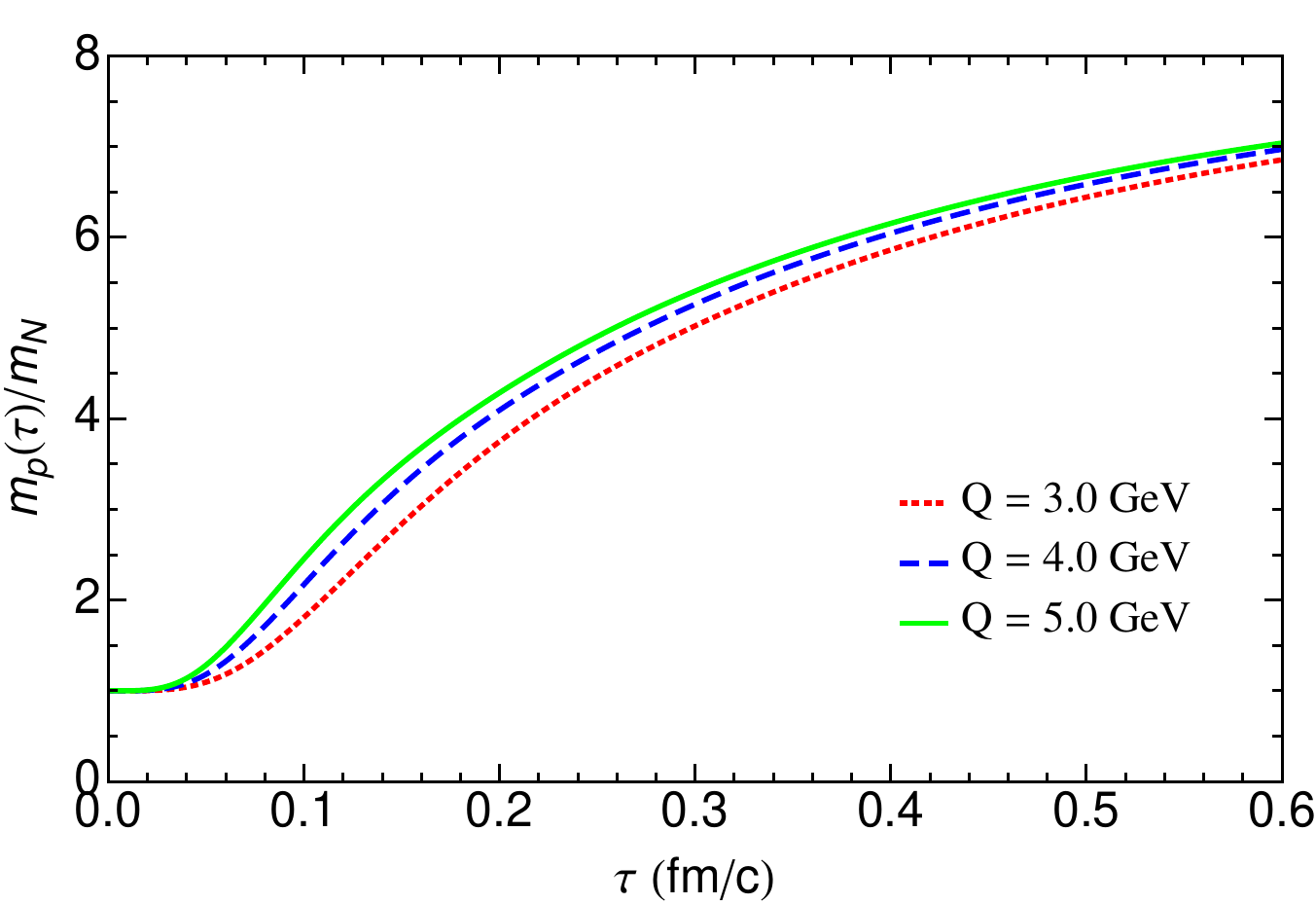}
\caption{\label{F2} Excitation energy per baryon in the central core of a Au projectile nucleus in the center-of-momentum frame for $\sqrt{s_{NN}} = 200$ GeV as a function of proper time.  The result is mildly sensitive to the choice of $Q$.}
\end{minipage} 
\end{figure}

\section{Densities, Entropies, Temperatures, and Chemical Potentials}

Both elementary considerations \cite{AKM1980} and detailed model calculations \cite{C1984,GC1986} show that immediately after the collision the projectile tube located at ${\bf r}_\perp$ is compressed by a factor of ${\rm e}^{\Delta y}$, where $\Delta y$ is its rapidity loss, and similarly for the target tube.  This is a boost invariant factor.  Since the trajectories of the tubes along the beam direction depend on ${\bf r}_\perp$ the receding projectile baryon fireball possesses a longitudinal shear.  How the baryon density distribution appears in the overall center of momentum frame requires careful consideration.  Figure 3 is a contour plot of the proper baryon density in the $\eta$-$r_\perp$ plane for central AuAu collisions at $\sqrt{s_{NN}}$ = 62.4 GeV.  It shows a characteristic swept wing pattern.  The central core suffers a greater rapidity loss and corresponding compression, whereas the periphery of the nuclear fireball suffers less rapidity loss and less compression.  At these high energies space-time rapidity $\eta$ is nearly equal to momentum-space rapidity $y$, hence the swept wing.  Figure 4 shows the contour plot at $\sqrt{s_{NN}}$ = 200 GeV. 
\begin{figure}[h]
\begin{minipage}{18pc}
\includegraphics[width=18pc]{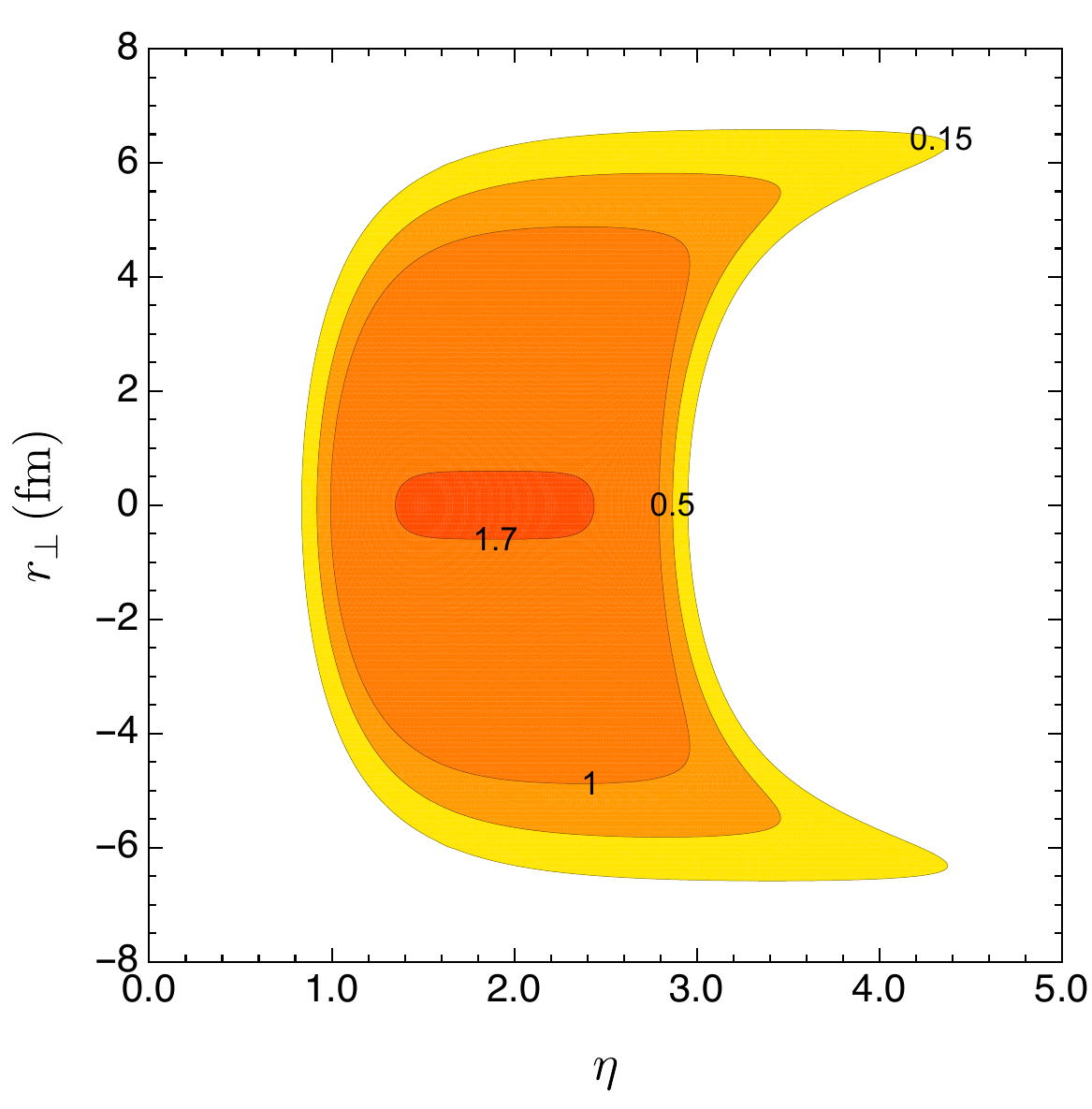}
\caption{Contour plot of the proper baryon density for central collisions of Au nuclei at $\sqrt{s_{NN}}$= 62.4 GeV.  The numbers are in units of baryons per cubic fm.}
\end{minipage}\hspace{2pc}%
\begin{minipage}{18pc}
\includegraphics[width=18pc]{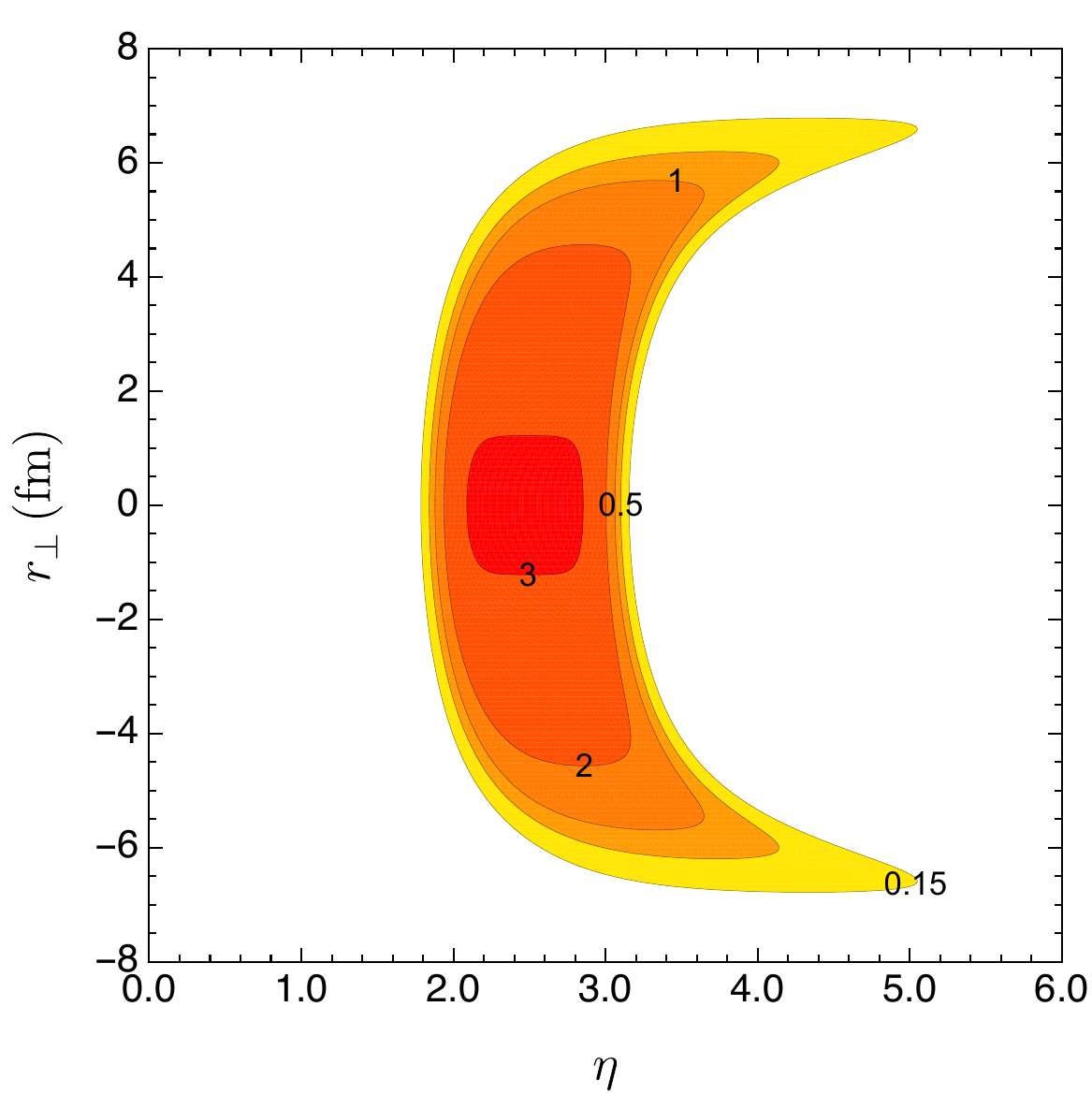}
\caption{Contour plot of the proper baryon density for central collisions of Au nuclei at $\sqrt{s_{NN}}$= 200 GeV.  The numbers are in units of baryons per cubic fm.}
\end{minipage} 
\end{figure}

\newpage

It is reasonable to assume that the matter in the baryonic fireballs will equilibrate on the same time scale as in the central region where the chemical potential is essentially zero.  It follows that one then needs an equation of state to obtain the temperature and chemical potential for given values of baryon and energy density.  For this we use a crossover equation of state that switches from a hadron resonance gas with excluded volumes to a perturbative QCD plasma and which accurately reproduces lattice results \cite{Albright}.  We use this equation of state to extrapolate to higher chemical potentials than so far reported in lattice simulations.  Adiabatic trajectories in the $T$-$\mu$ plane for three different values of ${\bf r}_\perp$ for central AuAu collisions at $\sqrt{s_{NN}}$= 200 GeV are shown in Fig. 5 along with the corresponding entropy per baryon ratios.  The curves are terminated at $T=120$ MeV because it is not known when chemical freeze-out would occur.  A plot of the entropy per baryon versus the rapidity of the streak or tube is shown in Fig. 6.  These values are precisely in the range where a suspected critical point may exist.  
\begin{figure}[h]
\begin{minipage}{18pc}
\includegraphics[width=18pc]{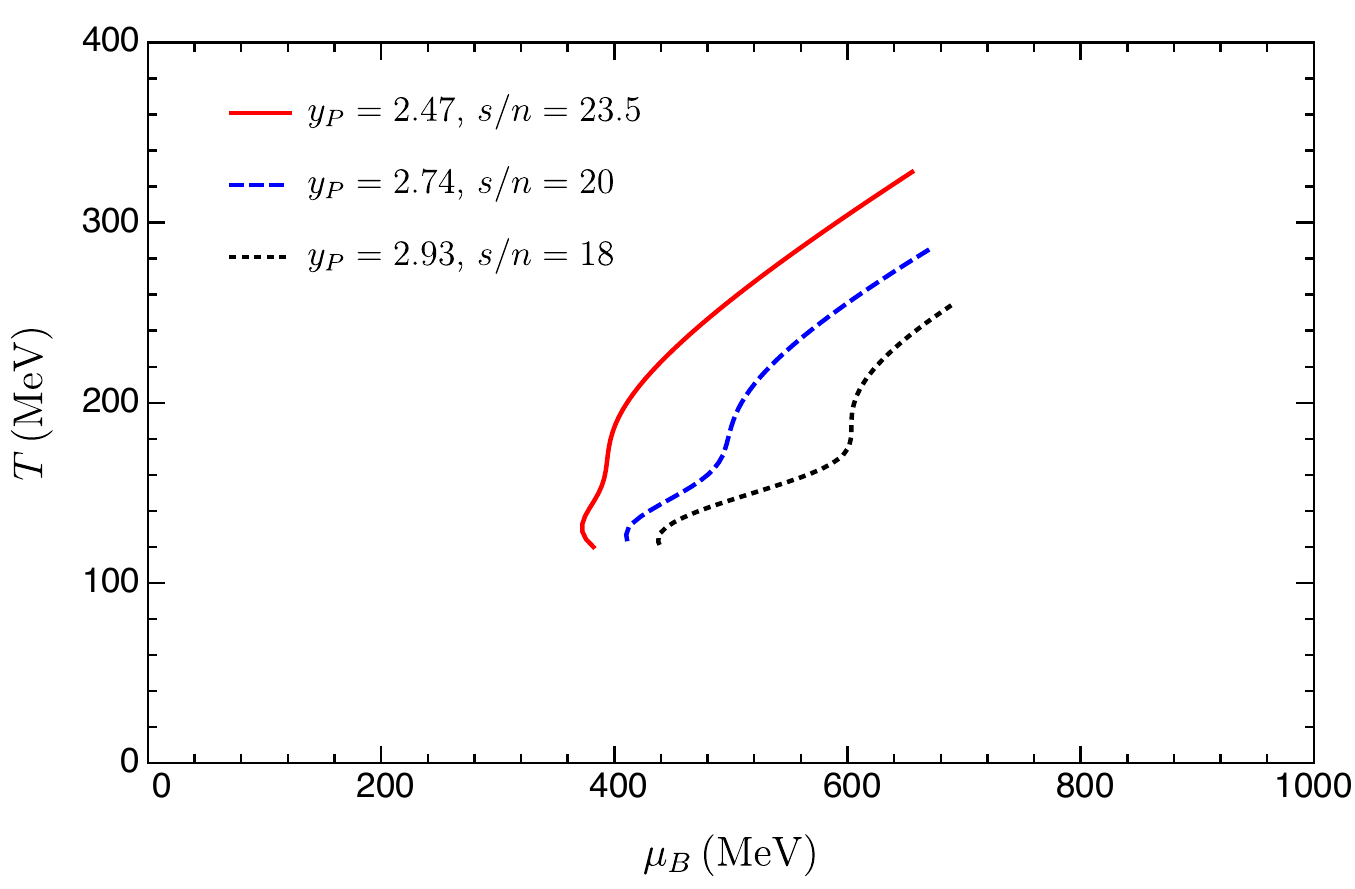}
\caption{Adiabats for three different projectile rapidities corresponding to different values of $r_\perp$ at $\sqrt{s_{NN}}$= 200 GeV.}
\end{minipage}\hspace{2pc}%
\begin{minipage}{18pc}
\includegraphics[width=18pc]{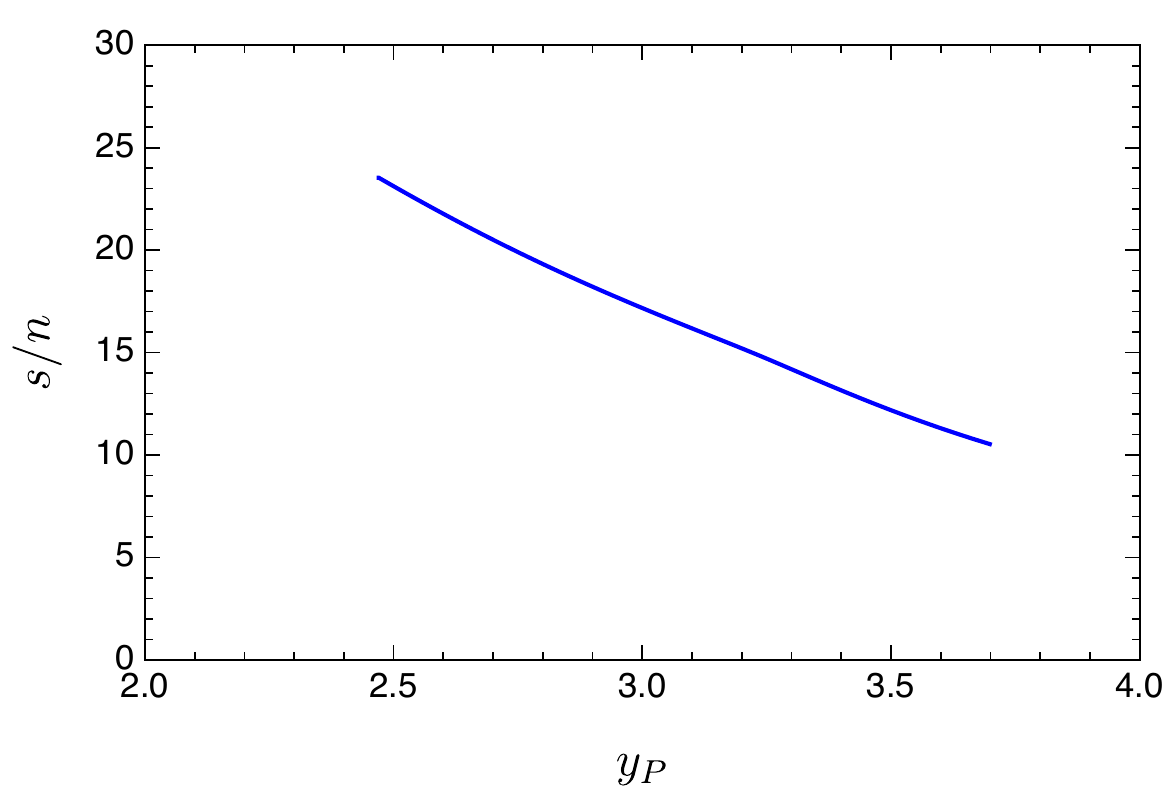}
\caption{Entropy per baryon as a function of rapidity corresponding to different values of $r_\perp$ at $\sqrt{s_{NN}}$= 200 GeV.  A scan through rapidity at fixed beam energy may reveal a critical point.}
\end{minipage} 
\end{figure}

\newpage

Both the baryon density and the entropy per baryon in the central core are increasing functions of beam energy.  For AuAu collisions at $\sqrt{s_{NN}}$= 62.4 GeV the rapidity of the central core is $y_P$ = 1.89 with $s/n$ = 16.3.  For PbPb collisions at the LHC the corresponding numbers for $\sqrt{s_{NN}}$= 2.76 (5.02) TeV are $y_P$ = 3.76 (4.06) with $s/n$ = 49.6 (58.2).  The entropy per baryon ratios at the LHC in collider mode are probably too high to be relevant for a search for a critical point.  However, if the LHC does run in fixed target mode then $\sqrt{s_{NN}}$= 72 GeV, which is relevant for the search.

\section{Conclusion}

We employed the McLerran-Venugopalan model to calculate the energy and rapidity loss and internal excitation of baryons in high energy heavy ion collisions.  Very similar results should be obtained in different pictures of heavy ion collisions.  We found that baryon densities in the fireballs outside the central rapidity region reach values an order of magnitude greater than normal nuclear matter.  Assuming equilibration on the same time scale as in the central rapidity region, we used a crossover equation of state to infer the temperatures, chemical potentials, and entropy densities.  For  $\sqrt{s_{NN}}$= 62.4 and 200 GeV at RHIC, and $\sqrt{s_{NN}}$= 72 GeV at the LHC in fixed target mode, the entropy per baryon spans a range on both sides of a suspected critical point in the QCD phase diagram.  Even a rapidity scan at fixed beam energy might be able to discern a critical point.  We have also studied the dependence on projectile and target size including asymmetric collisions, non-zero impact parameters, and even tip-tip collisions of UU collisions; these results will be reported elsewhere \cite{BigPaper,thesis}.  That needs to be followed up with detailed modeling using 3-dimensional relativistic viscous hydrodynamics with initial conditions as reported here.

We hope to have motivated further theoretical and experimental studies to probe the equation of state at the highest baryon densities achievable in a laboratory setting.\\

This work was supported by the U.S. Department of Energy Grant DE-FG02-87ER40328.  ML was also supported by a Doctoral Dissertation Fellowship from the University of Minnesota.






\begin{thebibliography}{00}


\bibitem{BJ1983}
J. D. Bjorken, Phys. Rev. D {\bf 27}, 140 (1983).

\bibitem{AKM1980}
R. Anishetty, P. Koehler and L. McLerran, Phys. Rev. D {\bf 22}, 2793 (1980).

\bibitem{LKRapid}
M. Li and J. I. Kapusta, Phys. Rev. C {\bf 95}, 011901(R) (2017).

\bibitem{McLerranVenu}
L. D. McLerran and R. Venugopalan, Phys. Rev.  D {\bf 49}, 2233 (1994); 3352 (1994).

\bibitem{CFKL2015}
G. Chen, R. J. Fries, J. I. Kapusta, and Y. Li, Phys. Rev. C {\bf 92}, 064912 (2015).

\bibitem{LK2016}
M. Li and J. I. Kapusta, Phys. Rev. C {\bf 94}, 024908 (2016).

\bibitem{MK2002}
I. N. Mishustin and J. I. Kapusta, Phys. Rev. Lett. {\bf 88}, 112501 (2002).

\bibitem{C1984}
L. P. Csernai, Phys. Rev. D {\bf 29}, 1945 (1984).

\bibitem{GC1986}
M. Gyulassy and L. P. Csernai, Nucl. Phys. {\bf A460}, 723 (1986).

\bibitem{Albright}
M. Albright, J. Kapusta, and C. Young, Phys. Rev. C {\bf 90}, 024915 (2014); {\bf 92}, 044904 (2015).

\bibitem{BigPaper}
M. Li and J. I. Kapusta, to appear.

\bibitem{thesis}
M. Li, Doctoral Dissertation, University of Minnesota, August 2018.


\end{thebibliography}



\end{document}